\documentclass{PoS}
\usepackage{bm}
\title{Flavour Anomalies}

\ShortTitle{Flavour Anomalies}

\author{\speaker{Antonio Pich} 
		\\
        IFIC, Universitat de Val\`encia -- CSIC\\
        Parc Cient\'{\i}fic, Catedr\'atico Jos\'e Beltr\'an 2, E-46980 Paterna, Spain\\
        E-mail: \email{Antonio.Pich@ific.uv.es}}

\abstract{The experimental data on $b\to c\tau\bar\nu_\tau$ and $b\to s\ell^+\ell^-$ transitions exhibit sizeable discrepancies with the Standard Model expectations. We present an overview of the present status and discuss possible interpretations within a model-independent effective Lagrangian approach. We also briefly elaborate on some other claimed flavour anomalies such as the recently observed CP asymmetry in $D^0$ decays or the $K^0\to \pi\pi$ ratio $\varepsilon_K'/\varepsilon^{\phantom{'}}_K$. The Standard Model prediction for the direct $CP$-violating ratio $\varepsilon_K'/\varepsilon^{\phantom{'}}_K$ agrees with its measured value, once all theoretical ingredients are correctly taken into account.
}

\FullConference{7th Annual Conference on Large Hadron Collider Physics - LHCP2019\\
		20-25 May, 2019\\
		Puebla, Mexico}

\begin{document}

\section{Introduction}

The Standard Model (SM) provides a very successful description of flavour-changing transitions in terms of the nine charged-fermion masses and the four CKM quark-mixing parameters (plus the analogous neutrino masses and mixings, when they are incorporated). However, a more fundamental understanding of flavour is still lacking. Obviously, any observed deviations from the SM predictions trigger a lot of interest as they could provide the missing hints to uncover the underlying flavour dynamics.

Many flavour-related `anomalies' have been reported along the years:
$W\to\tau\nu$, $\tau\to\pi K_S\nu$, $b\to c\tau\nu$, $b\to s\mu^+\mu^-$,
$(g-2)_{\mu,e}$, 
$\varepsilon'_K/\varepsilon^{\phantom{'}}_K$ ($K^0\to\pi\pi$), $\varepsilon^{\phantom{'}}_K$, 
$\Delta A_{CP}$ ($D^0\to K^+K^-/\pi^+\pi^-$), $V_{cd}$, $V_{ub}$, $V_{ud}$, $B\to\tau\nu$, \ldots
While some of them could indeed be true signals of new phenomena, others may just originate from statistical fluctuations, underestimated systematics or even a deficient SM prediction or measurement. For instance, the long-term discrepancy between exclusive and inclusive determinations of $V_{cd}$ seems nowadays close to get solved through a more careful treatment of form factor parameters and data extrapolations \cite{Bigi:2016mdz,Bigi:2017njr,Bigi:2017jbd,Grinstein:2017nlq,Bernlochner:2017jka,Bernlochner:2017xyx,Gambino:2019sif}. The resulting (slightly larger) value of $V_{cd}$ would also eliminate the suggested tensions in $\varepsilon^{\phantom{'}}_K$, while the claimed anomaly in $\varepsilon'_K/\varepsilon^{\phantom{'}}_K$ results from a poor theoretical treatment of the final-state $\pi\pi$ dynamics \cite{Gisbert:2017vvj}.

In fact, an easy common explanation for all anomalies does not exist, within appealing models of new physics (NP). Model builders just choose two or three anomalies, according to their preferences, in order to fit them together within the same theoretical framework.
In this situation, separate analyses of the different observables are perhaps more enlightening.
In the following, I will focus on those topics where there has been more activity during the last two years.

\section{The kaon CP-violating ratio $\mathbf{\varepsilon_K'/\varepsilon^{\protect\phantom{'}}_K}$}

A tiny difference between the $CP$-violating ratios $\eta_{nm}\equiv \mathcal{M}[K_L^0\to \pi^n\pi^m]/ \mathcal{M}[K_S^0\to \pi^n\pi^m]\approx\varepsilon^{\phantom{'}}_K \approx 2.2\times 10^{-3}\, \mathrm{e}^{i\pi/4}$, where $nm=+-,00$ denote the final pion charges, was first measured by the CERN NA31 experiment \cite{Burkhardt:1988yh} and later confirmed at the $7.2\sigma$ level with the full data samples of NA31, NA48 and the Fermilab experiments E731 and KTeV \cite{Tanabashi:2018oca}:
\begin{equation}\label{eq:epsp}
\mathrm{Re}(\varepsilon'_K/\varepsilon^{\phantom{'}}_K) = \frac{1}{3}\,\left(
1 - \left|\frac{\eta_{00}}{\eta_{+-}}\right|^2\right) = (16.6\pm 2.3)\times 10^{-4}\, .
\end{equation}
This important measurement established the presence of direct $CP$ violation
in the decay amplitudes, confirming that $CP$ violation is associated with a $\Delta S=1$ transition as predicted by the CKM mechanism.

The first next-to-leading-order theoretical predictions gave values of $\varepsilon_K'/\varepsilon^{\phantom{'}}_K$ one order of magnitude smaller than (\ref{eq:epsp}), but it was soon realised that they were missing the important role of the final pion dynamics \cite{Pallante:1999qf,Pallante:2000hk,Pallante:2001he}. Once long-distance contributions are properly taken into account, the theoretical SM prediction turns out to be in good agreement with the experimental value, although the uncertainties are unfortunately large \cite{Gisbert:2017vvj,Cirigliano:2019cpi}:
\begin{equation}
\mathrm{Re}(\varepsilon'_K/\varepsilon^{\phantom{'}}_K)_{\mathrm{SM}} = (14\pm 5)\times 10^{-4}\, .
\end{equation}

The underlying physics can be easily understood from the kaon data themselves. Owing to Bose symmetry, the two pions in the final state must be in a $I=0$ or $I=2$ configuration. In the absence of QCD corrections, the corresponding $K\to\pi\pi$ decay amplitudes $\mathcal{A}_I\equiv A_I\,\mathrm{e}^{i\delta_I}$
are predicted to differ only by a $\sqrt{2}$ factor. However, their measured ratio is 16 times larger than that (a truly spectacular enhancement generated by the strong forces):
\begin{equation}
\omega\equiv\mathrm{Re}(A_2)/\mathrm{Re}(A_0) \approx 1/22\, ,
\qquad\qquad
\delta_0-\delta_2\approx 45^\circ\, .
\end{equation}
Moreover, they exhibit a huge phase-shift difference that manifests the relevance of final-state interactions and, therefore, the presence of large absorptive contributions to the $K\to\pi\pi$ amplitudes, specially to the isoscalar one. Writing $\mathcal{A}_I = 
\mathrm{Dis} (\mathcal{A}_I) + i\, \mathrm{Abs} (\mathcal{A}_I)$
and neglecting the small $CP$-odd components, the measured $\pi\pi$ scattering phase shifts at $\sqrt{s}=m_K$ imply that 
%
\begin{equation}\label{epsp_abs}
\mathrm{Abs}(A_0)/\mathrm{Dis}(A_0) \approx 0.82\, ,
\qquad\qquad\quad 
\mathrm{Abs}(A_2)/\mathrm{Dis}(A_2) \approx - 0.15\, .
\end{equation}
%

The direct $CP$-violating effect involves the interference between the two isospin amplitudes,
\begin{equation}\label{epsp_th}
\mathrm{Re}(\varepsilon'_K/\varepsilon^{\phantom{'}}_K)\, =\, -\frac{\omega}{\sqrt{2}\, |\varepsilon^{\phantom{'}}_K|}\,\left[\frac{\mathrm{Im} A_0}{\mathrm{Re} A_0} -
\frac{\mathrm{Im} A_2}{\mathrm{Re} A_2}\right]
\, =\,
 -\frac{\omega_+}{\sqrt{2}\, |\varepsilon^{\phantom{'}}_K|}\,\left[\frac{\mathrm{Im} A_0^{(0)}}{\mathrm{Re} A_0^{(0)}}\,\left( 1 -\Omega_{\mathrm{eff}}\right) -
\frac{\mathrm{Im} A_2^{\mathrm{emp}}}{\mathrm{Re} A_2^{(0)}}\right] .
\end{equation}
It is suppressed by the small ratio $\omega$ and, moreover, it is very sensitive to isospin-breaking (IB) corrections \cite{Ecker:1999kr,Cirigliano:2003nn,Cirigliano:2003gt}, parametrized by $\Omega_{\mathrm{eff}}=0.11\pm0.09$ \cite{Cirigliano:2019cpi}, because small IB corrections to 
$A_0$ feed into the small amplitude $A_2$ enhanced by the large factor $1/\omega$. In the right-hand side of Eq.~(\ref{epsp_th}), the $(0)$ superscript indicates the isospin limit, $\omega_+ = \mathrm{Re}(A_2^+)/\mathrm{Re}(A_0)$ is directly extracted from $K^+\to\pi^+\pi^0$ and $A_2^{\mathrm{emp}}$ contains the electromagnetic-penguin contribution to $A_2$ (the remaining contributions are included in 
$\Omega_{\mathrm{eff}}$).

Claims for small SM values of $\varepsilon'_K/\varepsilon^{\phantom{'}}_K$ usually originate from perturbative calculations that are unable to generate the physical phase shifts, {\it i.e.}, they predict $\delta_I = 0$ and, therefore, $\mathrm{Abs} (\mathcal{A}_I) = 0$, failing completely to understand the empirical ratios (\ref{epsp_abs}). This unitarity pitfall implies also incorrect predictions for the dispersive components, since they are related by analyticity with the absorptive parts: a large absorptive contribution generates a large dispersive correction that is obviously missed in those calculations. This perturbative problem is more severe in $\varepsilon'_K/\varepsilon^{\phantom{'}}_K$ because Eq.~(\ref{epsp_th}) involves a delicate numerical balance among the three contributing terms, and naive predictions sit precisely on a nearly-exact cancellation (a 40\% positive correction to the first term enhances the whole result by one order of magnitude).

The $\varepsilon'_K/\varepsilon^{\phantom{'}}_K$ anomaly was recently resurrected by the lattice RBC-UKQCD collaboration that reported 
$\mathrm{Re}(\varepsilon'_K/\varepsilon^{\phantom{'}}_K) = (1.38\pm 5.15\pm 4.59)\times 10^{-4}$ \cite{Bai:2015nea,Blum:2015ywa}. The uncertainties are still large, but the quite low central value implies a $2.1\sigma$ deviation from the experimental measurement. This has triggered a revival of the old naive estimates \cite{Buras:2015xba,Buras:2016fys}, some of them making also use of the lattice data \cite{Buras:2015yba,Kitahara:2016nld}, and a large amount of NP explanations (a list of references is given in Refs.~\cite{Gisbert:2017vvj,Cirigliano:2019cpi}). However, it is premature to derive physics implications from the current lattice simulations, since they are still unable to reproduce the known phase shifts. While the lattice determination of $\delta_2$ is only $1\sigma$ away from its physical value, $\delta_0$ disagrees with the experimental result by $2.9\sigma$, a much larger discrepancy that the one quoted for $\varepsilon'_K/\varepsilon^{\phantom{'}}_K$. Obviously, nobody suggests a NP contribution to the $\pi\pi$ elastic scattering phase shifts. 
The RBC-UKQCD collaboration is actively working in order to improve the present situation.

\section{Direct CP violation in charm}
 
\begin{figure}[tbh]
\begin{center}
\includegraphics[width=10.cm]{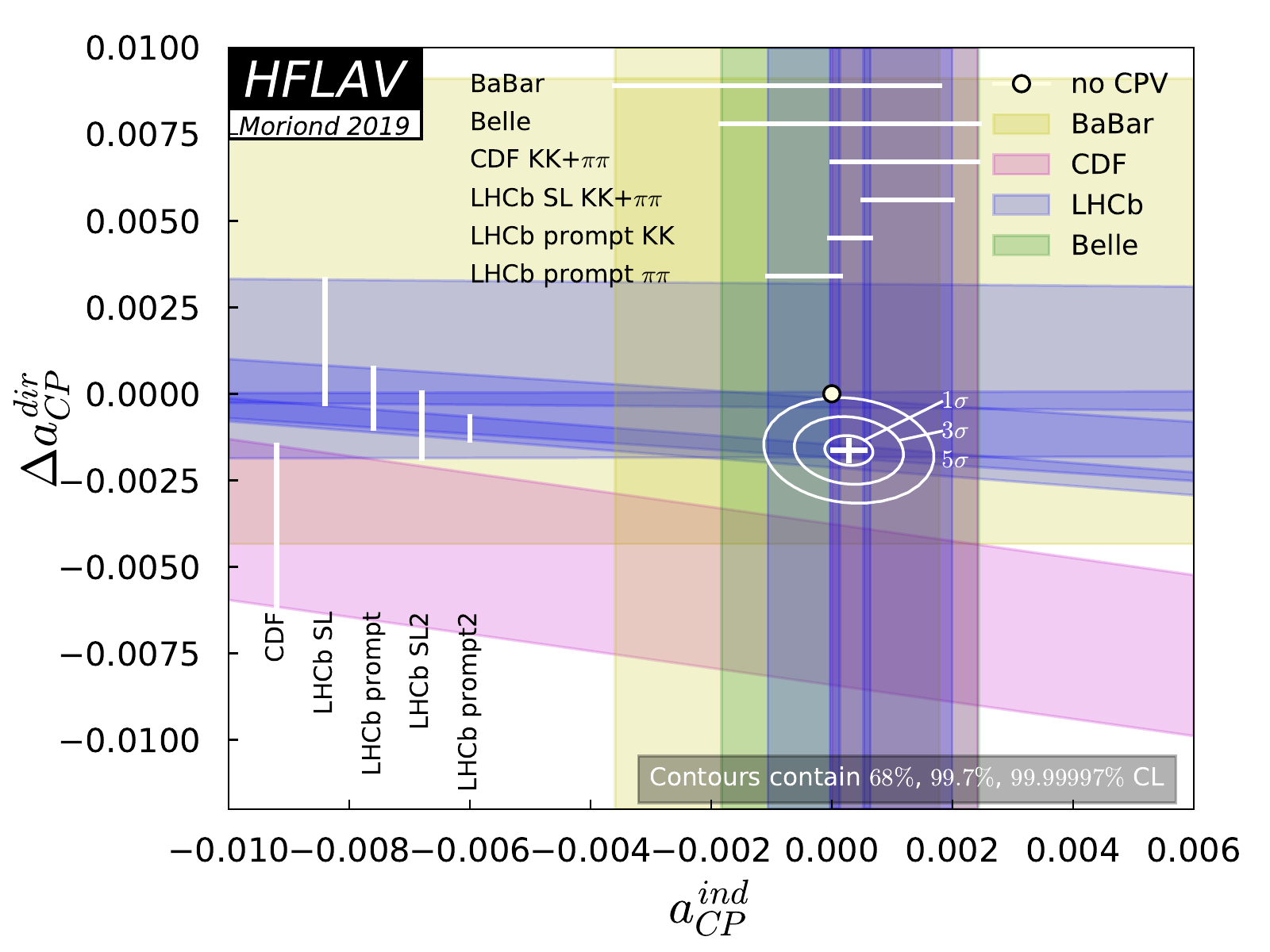}
\caption{World data on the direct and indirect CP-violating contributions
$\Delta a^{\mathrm{dir}}_{CP}$ and $a^{\mathrm{ind}}_{CP}$
 \protect\cite{Amhis:2019ckw}. The cross indicates the best fit values with their $1\sigma$ uncertainties, and the ellipses the two-dimensional $1\sigma$, $3\sigma$ and $5\sigma$ regions. The filled circle shows the point $(0,0)$, where the two CP-violating amplitudes vanish.}
\label{fig:D-CP}
\end{center}
\end{figure}

The first observation of CP violation in decays of charm hadrons has been recently reported by the LHCb collaboration, which has measured the difference between the CP asymmetries in $D^0\to K^+K^-$ and $D^0\to \pi^+\pi^-$~\cite{Aaij:2019kcg}:
\begin{equation}
\Delta A_{CP}\,\equiv A_{CP}(K^+K^-) - A_{CP}(\pi^+\pi^-)\, =\, 
(-15.4\pm 2.9)\times 10^{-4}\, ,
\end{equation}
where $A_{CP}(f)\equiv [\Gamma(D^0\to f)-\Gamma(\bar D^0\to f)]/[\Gamma(D^0\to f)+\Gamma(\bar D^0\to f)]$.
The measured time-integrated asymmetries can be decomposed in a direct contribution $a^{\mathrm{dir}}_{CP}(f)$ from CP violation in the decay amplitude and another component associated to CP violation either in $D^0$-$\bar D^0$ mixing or in the interference between mixing and decay. These two contributions can be disentangled through measurements of the time-decay distribution of the data sample.
%
The LHCb analysis concludes that, as expected, the CP signal comes primarily from direct CP violation: $\Delta a^{\mathrm{dir}}_{CP} = 
(-15.7\pm 2.9)\times 10^{-4}$~\cite{Aaij:2019kcg}. The HFLAV combination with previous measurements, shown in Figure~\ref{fig:D-CP}, gives \cite{Amhis:2019ckw}
\begin{equation}\label{eq:a_CP_dir}
\Delta a^{\mathrm{dir}}_{CP}\, =\, (-16.4\pm 2.8)\times 10^{-4}\, .
\end{equation}
%

The size of this asymmetry is larger than the naive SM expectation $|\Delta a^{\mathrm{dir}}_{CP}|\le 3\times 10^{-4}$ \cite{Chala:2019fdb,Khodjamirian:2017zdu}, based on perturbative QCD or light-cone sum rules estimates. However, the measured value could be explained within the SM with non-perturbative re-scattering effects of moderate size that enhance the U-spin invariant decay amplitude \cite{Grossman:2019xcj}. A direct CP asymmetry necessarily involves a non-zero strong phase-shift difference between (at least) two interfering amplitudes. Eq.~(\ref{eq:a_CP_dir}) indicates the presence of a large phase-shift, which can obviously not be reproduced through perturbative calculations. A reliable SM prediction of $\Delta a^{\mathrm{dir}}_{CP}$, including re-scattering corrections \cite{Buccella:2019kpn,Li:2019hho,Soni:2019xko,Cheng:2019ggx}, remains an interesting theoretical challenge.

\section{$\mathbf{b \boldsymbol{\to} c \tau\bar\nu_\tau}$ transitions}

\begin{figure}[tbh]
\begin{center}
\includegraphics[width=8.cm]{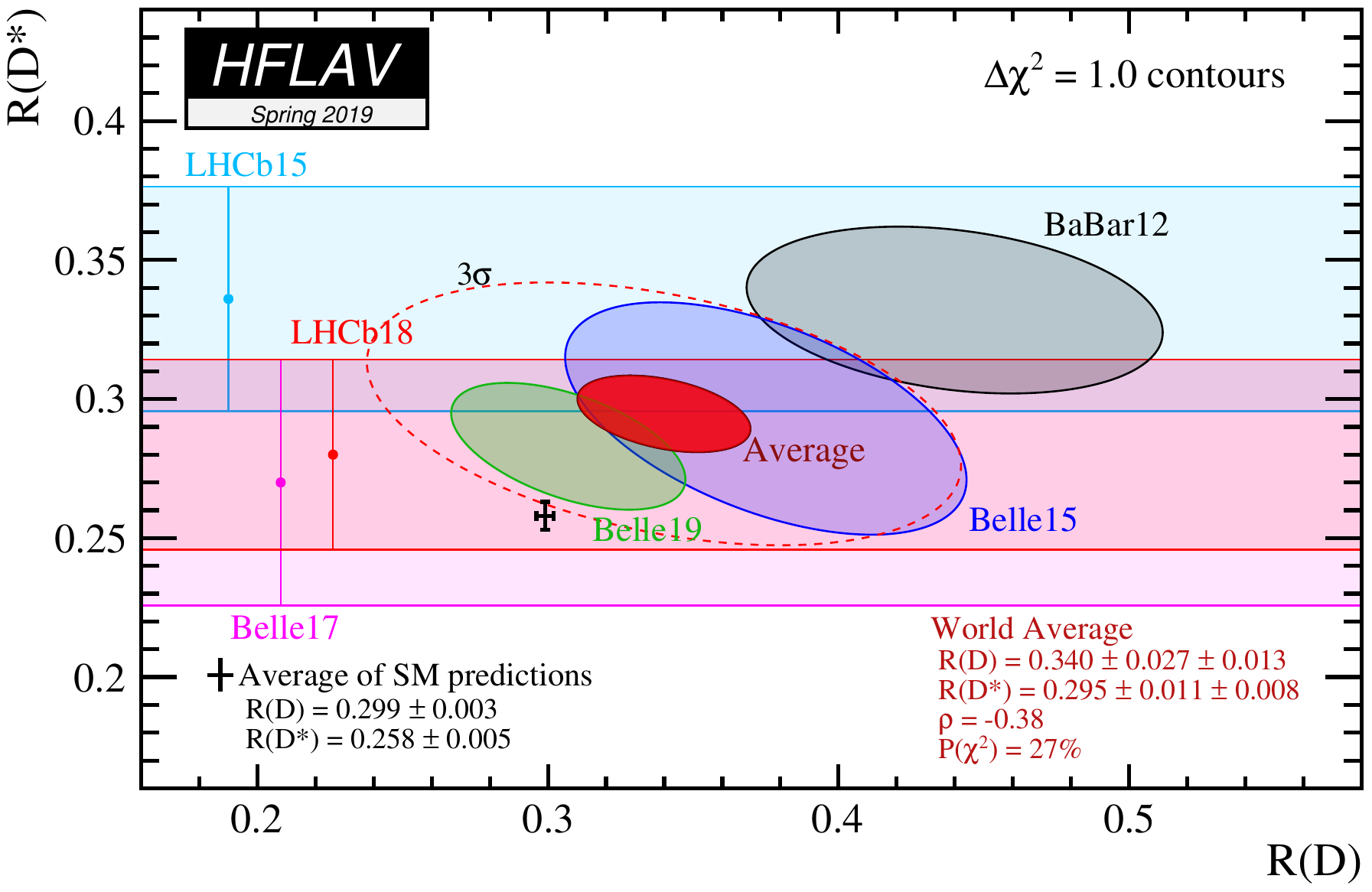}
\caption{$R(D)$ and $R(D^*)$ measurements, their world average (red ellipse) and SM predictions (cross)~\cite{Amhis:2019ckw}.}
\label{fig:RDRDs}
\end{center}
\end{figure}

Sizeable deviations from their predicted SM values have been found in the ratios
\begin{equation}
R(D^{(*)})\,\equiv\, \frac{\mathrm{Br}(\bar B\to D^{(*)} \tau^-\bar\nu_\tau)}{\mathrm{Br}(\bar B\to D^{(*)} \ell^-\bar\nu_\ell)}\, ,
\end{equation}
with $\ell = e,\mu$. The experimental averages shown in Figure~\ref{fig:RDRDs} exhibit a $3.1\sigma$ discrepancy with the SM predictions quoted by the HFLAV group~\cite{Amhis:2019ckw}, which increases to $3.7\sigma$ with the more updated theoretical values given in Ref.~\cite{Murgui:2019czp}:
\begin{equation}
R(D)^{\mathrm{SM}}\, =\, 0.300\,{}^{+\, 0.005}_{-\, 0.004}\, ,
\qquad\qquad\qquad
R(D^*)_{\mathrm{SM}}\, =\, 0.251\,{}^{+\, 0.004}_{-\, 0.003}\, .
\end{equation}
A large part of the hadronic form factor uncertainties cancels in these ratios, which are also independent of $V_{cb}$. Therefore, the measurements suggest a rather large violation of lepton-flavour universality that is quite unexpected in a tree-level SM transition. A similar $1.7\sigma$ deviation has been also found for the analogous $R(J/\psi)$ ratio of semileptonic $B_c\to J/\psi$ transitions \cite{Aaij:2017tyk}. Moreover, the measured longitudinal polarization of the $D^{*-}$ meson in $B^0\to D^{*-}\tau^+\nu_\tau$
\cite{Abdesselam:2019wbt} also differs from its SM value by $1.6\sigma$.

From the SM inclusive prediction $\mathrm{Br}(B\to X_c\tau\nu)/\mathrm{Br}(B\to X_c e\nu)= (0.222\pm 0.007)$ \cite{Freytsis:2015qca,Celis:2016azn}, which does not involve any form factors, and $\mathrm{Br}(B\to X_c\ell\nu) = (10.65\pm 0.16)\%$ \cite{Tanabashi:2018oca}, one finds that 
$\mathrm{Br}(B\to X_c\tau\nu) = (2.36\pm 0.08)\%$, in agreement with the LEP result $\mathrm{Br}(b\to X_c\tau\nu) = (2.41\pm 0.23)\%$ \cite{Tanabashi:2018oca}. Since the measured $R(D^{(*)})$ ratios imply 
$\mathrm{Br}(B\to D\tau\nu) + \mathrm{Br}(B\to D^*\tau\nu) = (2.33\pm 0.11)\%$, these two final states saturate the inclusive width and there is no space left for other decay modes, such as the $D^{**}$, that are expected to contribute more than $0.5\%$ \cite{Freytsis:2015qca}. Thus, there is also tension with the SM at the inclusive level.

The anomaly is dominated by the 2012--2013 BaBar results \cite{Lees:2012xj,Lees:2013uzd}, while the most recent Belle \cite{Huschle:2015rga,Hirose:2016wfn,Abdesselam:2019dgh} and LHCb \cite{Aaij:2015yra,Aaij:2017uff,Aaij:2017deq} measurements are closer to the SM values. 
Moreover, the normalized $q^2$ distributions measured by BaBar \cite{Lees:2013uzd} and Belle \cite{Huschle:2015rga} do not favour large deviations from the SM \cite{Celis:2016azn}. One must also take into account that the needed enhancement of the $b\to c\tau \nu$ transition is constrained by the cross-channel $b\bar c\to\tau \nu$. A conservative (more stringent) upper bound $\mathrm{Br}(B_c\to\tau\nu) < 30\%$ (10\%) can be extracted from the $B_c$ lifetime \cite{Celis:2016azn,Alonso:2016oyd} (LEP data \cite{Akeroyd:2017mhr}).

Taking the available experimental information at face value, one can investigate the possible types of underlying NP interactions with a generic low-energy effective Hamiltonian,
\begin{equation}
{\cal H}_{\mathrm{eff}}^{b\to c \ell \nu}\, =\, \frac{4G_F}{\sqrt{2}}\, V_{cb}\,\left\{ \left(  1 + C_{V_L} \right) {\cal O}_{V_L} + C_{V_R} {\cal O}_{V_R}
+ C_{S_R} {\cal O}_{S_R} + C_{S_L} {\cal O}_{S_L} + C_{T} {\cal O}_T \right\}  + \mathrm{h.c.}
\end{equation}
where
\begin{equation}
{\cal O}_{V_{L,R}} = \left( \bar{c}\, \gamma^{\mu} b_{L,R} \right)\left( \bar{\ell}_L \gamma_{\mu} \nu_{\ell L} \right)\, ,
\quad\;  
{\cal O}_{S_{L,R}} = \left( \bar{c}\,  b_{L,R} \right)\left( \bar{\ell}_R \nu_{\ell L} \right)\, ,
\quad\;
{\cal O}_{T} = \left( \bar{c}\, \sigma^{\mu \nu} b_L \right)\left( \bar{\ell}_R \sigma_{\mu \nu} \nu_{\ell L} \right)\, .
\end{equation}
The SM corresponds to $C_i=0$. Since potential NP contributions to the light-lepton couplings are highly constrained by $b\to c (e,\mu)\nu$ data~\cite{Jung:2018lfu}, one can safely consider that NP effects are only present for the $\tau$.
A global fit to all available experimental information has been recently done in Ref.~\cite{Murgui:2019czp},\footnote{An extensive list of references to previous analyses, most of them performed with a single mediator or operator and with partial data information, can be found in Ref.~\protect\cite{Murgui:2019czp}.}
neglecting CP-violating contributions ({\it i.e.}, with real $C_i$) and taking $C_{V_R} = 0$. The second condition follows from the assumption that the electroweak symmetry breaking is linearly realized at the electroweak scale, which implies that $C_{V_R}$ is flavour independent \cite{Cata:2015lta}. The fitted results clearly indicate that NP contributions are needed (much lower $\chi^2$ than in the SM), but they do not show any strong preference for a particular Wilson coefficient ($\chi^2_{\mathrm{min}}/\mathrm{d.o.f.} = 37.4/54$) \cite{Murgui:2019czp}:
\begin{equation}
C_{V_L} = 0.09\,{}^{+\, 0.13}_{-\, 0.12}\, ,
\qquad
C_{S_R} = 0.09\,{}^{+\, 0.12}_{-\, 0.61}\, ,
\qquad
C_{S_L} = -0.14\,{}^{+\, 0.52}_{-\, 0.07}\, ,
\qquad
C_{T} = 0.008\,{}^{+\, 0.046}_{-\, 0.044}\, .
\end{equation}
While none of the fitted coefficients are required to be non-zero, the simplest interpretation of this solution is a global modification of the SM. In fact, setting all coefficients but $C_{V_L}$ to zero one also gets a good fit. In addition to this SM-like global minimum, a second local minimum ($\chi^2_{\mathrm{min}}/\mathrm{d.o.f.} = 40.4/54$) is found with larger non-SM contributions~\cite{Murgui:2019czp}:
\begin{equation}
C_{V_L} = 0.34\,{}^{+\, 0.05}_{-\, 0.07}\, ,
\qquad
C_{S_R} = -1.10\,{}^{+\, 0.48}_{-\, 0.07}\, ,
\qquad
C_{S_L} = -0.30\,{}^{+\, 0.11}_{-\, 0.50}\, ,
\qquad
C_{T} = 0.093\,{}^{+\, 0.029}_{-\, 0.030}\, .
\end{equation}

The measured $D^*$ longitudinal polarization fraction $F_L^{D^*}$ has a strong impact on the analysis because, with the four fitted operators, its predicted value remains always  below the $1\sigma$ experimental region. Including $C_{V_R}$ in the fit helps to remove the tension with the $B\to D^*$ data and opens new (not satisfactory) fine-tuned solutions where the SM coefficient becomes very small, its effect being substituted by several sizeable NP contributions, especially $C_{V_R}$.
More precise experimental data is needed to clarify the current situation. 
If the $b\to c$ anomaly remains, an improved measurement of $F_L^{D^*}$ could have major implications in its theoretical interpretation.

\section{$\mathbf{b \boldsymbol{\to} s\, \bm{\ell} \bm{\ell}}$ transitions}

\begin{figure}[tbh]
\begin{center}
\includegraphics[width=8cm]{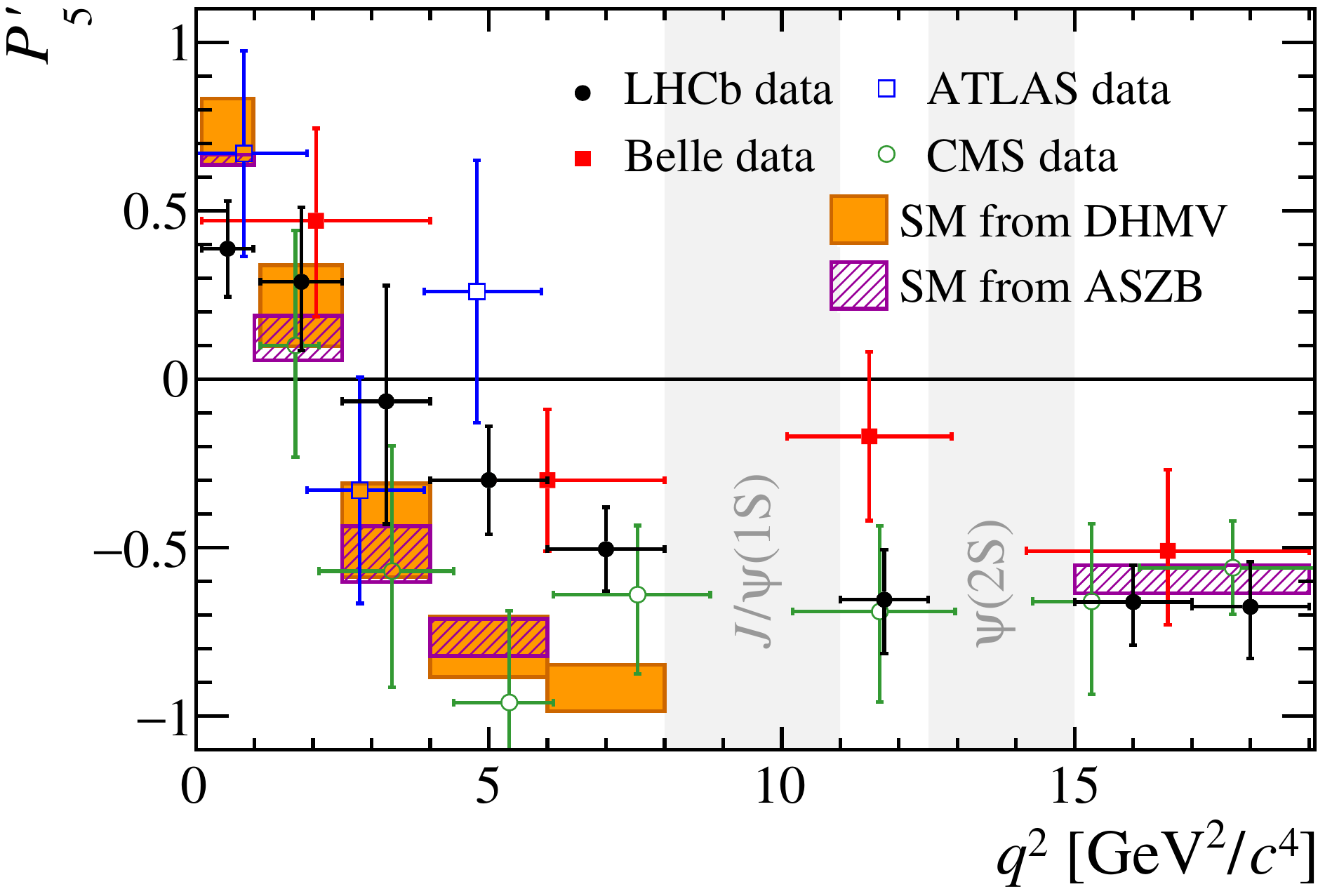}
\caption{Comparison between the predicted SM values of $P'_5$ and the experimental measurements \protect\cite{Dettori:2018igw}.}
\label{fig:Pp5}
\end{center}
\end{figure}

Several $b\to s\mu^+\mu^-$ rates have been found at LHCb to be consistently lower than their SM predictions: $B^+\to K^+\mu^+\mu^-$ \cite{Aaij:2014pli,Aaij:2012vr}, $B^+\to K^{*+}\mu^+\mu^-$ \cite{Aaij:2014pli}, $B^0_d\to K^0\mu^+\mu^-$ \cite{Aaij:2014pli}, $B^0_d\to K^{*0}\mu^+\mu^-$ \cite{Aaij:2016flj,Aaij:2015oid,Aaij:2013iag,Aaij:2013qta}, $B^0_s\to \phi\mu^+\mu^-$ \cite{Aaij:2015esa,Aaij:2013aln} and $\Lambda_b^0\to\Lambda\mu^+\mu^-$ \cite{Aaij:2015xza,Aaij:2013mna}.  
The angular and invariant-mass distributions of the final decay products in $B\to K^{*}\mu^+\mu^-$ have been also studied by ATLAS \cite{Aaboud:2018krd}, BaBar \cite{Lees:2015ymt}, Belle \cite{Wehle:2016yoi,Wei:2009zv,Abdesselam:2016llu}, CDF \cite{Aaltonen:2011ja}, CMS \cite{Sirunyan:2017dhj,Khachatryan:2015isa} and LHCb \cite{Aaij:2016flj,Aaij:2015oid,Aaij:2013iag,Aaij:2013qta}. The four-body $K\pi\mu^+\mu^-$ final state
provides a rich variety of angular dependences, making possible to disentangle 
the different dynamical contributions. Particular attention has been devoted to specific combinations of angular observables that are free from form-factor uncertainties in the heavy-quark mass limit, the so called optimized observables $P'_i(q^2)$ \cite{DescotesGenon:2012zf}, where $q^2$ is the dilepton invariant-mass squared. A sizeable discrepancy with the SM prediction \cite{Descotes-Genon:2014uoa,Altmannshofer:2014rta,Straub:2015ica}, shown in Figure~\ref{fig:Pp5}, has been identified in two adjacent bins of the $P'_5$ distribution, just below the $J/\psi$ peak.
Belle has also analyzed $K^{*} e^+e^-$ final states \cite{Wehle:2016yoi,Wei:2009zv,Abdesselam:2016llu}, finding them compatible with the SM expectations.

The SM predictions for the previous observables suffer from hadronic uncertainties that are not easy to quantify. However, LHCb has also reported sizeable violations of lepton universality, at the $2.1$-$2.5\,\sigma$ level, through the ratios \cite{Aaij:2017vbb}
\begin{equation}
R_{K^{*0}}\,\equiv\,\frac{\Gamma(B^0_d\to K^{*0}\mu^+\mu^-)}{\Gamma(B^0_d\to K^{*0} e^+e^-)}\, =\, 
\left\{ \begin{array}{lc} 
0.66\,{}^{+\, 0.11}_{-\, 0.07}\pm 0.03\, ,\qquad & 
q^2\in [0.045,1.1]\;\mathrm{GeV}^2\, ,
\\[5pt]
0.69\,{}^{+\, 0.11}_{-\, 0.07}\pm 0.05\, ,\qquad &
q^2\in [1.1,6.0]\;\mathrm{GeV}^2\, ,
\end{array}\right.
\end{equation}
and \cite{Aaij:2019wad}
\begin{equation}
R_{K}\,\equiv\,\left.\frac{\Gamma(B^+\to K^+\mu^+\mu^-)}{\Gamma(B^+\to K^+ e^+e^-)}\right|_{q^2\in[1.1,6.0]\;\mathrm{GeV}^2}\, =\, 0.846\,{}^{+\, 0.060}_{-\, 0.054}\,{}^{+\, 0.016}_{-\, 0.014}\, ,
\end{equation}
which constitute very clean probes of NP contributions. Owing to their larger uncertainties, the recent Belle measurements of $R_{K^{*}}$ \cite{Abdesselam:2019wac} and $R_{K}$ \cite{Abdesselam:2019lab} are compatible with the SM as well as with LHCb.

Global fits to the $b\to s\ell^+\ell^-$ data with an effective low-energy Lagrangian
\begin{equation}
{\cal L}_{\mathrm{eff}}\, =\, \frac{G_F}{\sqrt{2}}\, V_{td}^{\phantom{*}} V_{ts}^*\,\frac{\alpha}{\pi}\;\sum_{i,\ell} C_{i,\ell}\, O_i^\ell
\end{equation}
show a clear preference for NP contributions to the operators
$O_9^\ell = (\bar s_L\gamma_\mu b_L) (\bar \ell\gamma^\mu\ell)$ and 
$O_{10}^\ell = (\bar s_L\gamma_\mu b_L) (\bar \ell\gamma^\mu\gamma_5\ell)$, with $\ell=\mu$ \cite{Aebischer:2019mlg,Alguero:2019ptt,Ciuchini:2019usw,Datta:2019zca,Kowalska:2019ley,Arbey:2019duh,Alok:2019ufo}. Although the different analyses tend to favour slightly different solutions, two main common scenarios stand out: either $\delta C_{9,\mu}^{\mathrm{NP}}\approx -0.98$ or
$\delta C_{9,\mu}^{\mathrm{NP}} = - \delta C_{10,\mu}^{\mathrm{NP}}\approx -0.46$. Both constitute large shifts ($-24\%$ and $-11\%$, respectively) from the SM values: $C_{9,\mu}^{\mathrm{SM}}(\mu_b) \approx 4.1$ and $C_{10,\mu}^{\mathrm{SM}}(\mu_b) \approx -4.3$, at $\mu_b = 4.8$~GeV. The first possibility is slightly preferred by the global analysis of all data, while the left-handed NP solution accommodates better the lepton-flavour-universality-violating observables \cite{Alguero:2019ptt}.

The left-handed scenario is theoretically appealing because it can be easily generated through $SU(2)_L\otimes U(1)_Y$-invariant effective operators at the electroweak scale that, moreover, could also provide an explanation to the $b\to c\tau\nu$ anomaly. This possibility emerges naturally from the so-called $U_1$ vector leptoquark model \cite{Angelescu:2018tyl}, and can be tested experimentally, since it implies a $b\to s\tau^+\tau^-$ rate three orders of magnitude larger than the SM expectation \cite{Capdevila:2017iqn}. For a recent review of theoretical models with a quite complete list of references, see Ref.~\cite{Bifani:2018zmi}.

\section{Summary}

Uncovering the fundamental dynamics behind flavour-changing transitions and CP-violating phenomena is one of the main pending questions in particle physics. In the SM, flavour emerges from the Yukawa interactions with the scalar Higgs doublet, the less understood part of the electroweak Lagrangian that is more open to theoretical speculations.

Sizeable deviations from the SM expectations have been identified in $b \to c \tau\bar\nu$ and $b\to s \ell\ell$ data. Whether they represent the first signals of new phenomena or just result from statistical fluctuations and/or underestimated systematics remains to be understood. New experimental input from LHC and Belle-II should soon clarify the situation.
A confirmation of the current flavour anomalies
would constitute clear evidence of NP interactions and, moreover, would allow us to infer their low-energy structure, providing precious hints on the underlying dynamics.

\section*{Acknowledgements}

I want to thank the organizers of LHCP2019 for the invitation to present this overview. I also thank
V. Cirigliano, H. Gisbert, M. Jung, C. Murgui, A. Pe\~nuelas and A. Rodr\'{\i}guez-S\'anchez for a very productive and enjoyable collaboration.
This work has been supported in part by the Spanish Government and ERDF funds from the EU Commission [grant FPA2017-84445-P], the Generalitat Valenciana [grant Prometeo/2017/053] and the Spanish Centro de Excelencia Severo Ochoa Programme [grant SEV-2014-0398].

\bibliographystyle{JHEP}
\bibliography{FlavAnom}

%

\end{document}